\documentclass{article}

\usepackage{microtype}
\usepackage{graphicx}
\usepackage{subfigure}
\usepackage{booktabs}
\usepackage{rotating}
\usepackage{hyperref}

\usepackage[accepted]{ml4astro2025}

\usepackage{amsmath}
\usepackage{amssymb}
\usepackage{mathtools}
\usepackage{amsthm}

\usepackage[capitalize,noabbrev]{cleveref}

\theoremstyle{plain}

\theoremstyle{definition}

\theoremstyle{remark}

\usepackage[textsize=tiny]{todonotes}
\newcommand{\R}{\mathbb{R}}

\mlforastrotitlerunning{MMVAE for Type Ia Supernovae}

\begin{document}

\twocolumn[
\mlforastrotitle{Mixture-of-Expert Variational Autoencoders for Cross-Modality Embedding of Type~Ia Supernova Data}

\begin{mlforastroauthorlist}
\mlforastroauthor{Yunyi Shen*}{yyy,xxx}
\mlforastroauthor{Alexander T. Gagliano*}{yyy,aaa,comp}
\end{mlforastroauthorlist}
\mlforastroaffiliation{xxx}{Department of Electrical Engineering and Computer Science, Massachusetts Institute of Technology, Cambridge, MA 02139, USA}
\mlforastroaffiliation{aaa}{Department of Physics, Massachusetts Institute of Technology, Cambridge, MA 02139, USA}
\mlforastroaffiliation{yyy}{The NSF Institute for AI and Fundamental Interactions}
\mlforastroaffiliation{comp}{Center for Astrophysics | Harvard \& Smithsonian, 60 Garden Street, Cambridge, MA 02138, USA}

\mlforastrocorrespondingauthor{Yunyi Shen}{yshen99@mit.edu}
\mlforastrocorrespondingauthor{Alexander Gagliano}{gaglian2@mit.edu}

\mlforastrokeywords{Machine Learning, ICML}

\vskip 0.3in
]

\printAffiliationsAndNotice{\mlforastroEqualContribution}

\begin{abstract}
Time-domain astrophysics relies on heterogeneous and multi-modal data. Specialized models are often constructed to extract information from a single modality, but this approach ignores the wealth of cross-modality information that may be relevant for the tasks to which the model is applied. In this work, we propose a multi-modal, mixture-of-expert variational autoencoder to learn a joint embedding for supernova light curves and spectra. Our method, which is inspired by the perceiver architecture, natively accommodates variable-length inputs and the irregular temporal sampling inherent to supernova light curves. We train our model on radiative transfer simulations and validate its performance on cross-modality reconstruction of supernova spectra and physical parameters from the simulation. Our model achieves superior performance in cross-modality generation to nearest-neighbor searches in a contrastively-trained latent space, showing its promise for constructing informative latent representations of multi-modal astronomical datasets. 
\end{abstract}

\section{Introduction}
\textbf{Diversity of Type Ia Supernovae:} Type Ia supernovae (SNe~Ia) are the thermonuclear explosions of white dwarfs in binary systems. These explosions occur as the primary star accretes mass from its companion and approaches the Chandrasekhar limit \citep{hoyle1960nucleosynthesis, liu2023type}. Because of the relatively robust correlations among their observational properties \citep[e.g., the ``Phillips relation" between an explosion's maximum luminosity and its dimming rate; ][]{pskovskii1977light, phillips1993absolute, phillips1999reddening}, SNe~Ia have increasingly been used as standardizable candles to measure cosmological distances (e.g., \citealt{riess1998observational,perlmutter1999measurements,DES2024SNeIa}).

As more SNe~Ia are observed from wide-field imaging surveys (such as the Zwicky Transient Facility; \citealt{bellm2018zwicky}), however, researchers have discovered SNe~Ia with atypical light curve features (e.g., early flux excesses; \citealt{wang2024bee}). The spectra of these SNe, which measure their wavelength-specific emission, have revealed important differences between the temperatures, velocities, and compositions of these more peculiar explosions and their normal counterparts. Understanding the \textit{correlations} within and across these data modalities will allow astronomers to study the physics underlying this diversity and automatically identify unusual SNe~Ia in upcoming surveys.

\textbf{Legacy Survey of Space and Time:} With the upcoming Vera C. Rubin Observatory Legacy Survey of Space and Time \citep[LSST;][]{ivezic2019lsst} imaging the full Southern Sky every 3-4 days for a decade, we expect to obtain light curves for $\sim$1~M SNe each year. Our spectroscopic resources will only be able to observe the smallest subset of these discovered explosions (at most 0.1-1\%). As a result, representation learning techniques applied to these datasets must be able to accommodate missing modalities at inference time. Methods able to extract multi-modal correlations from these highly imbalanced datasets will play a powerful role in helping astronomers prioritize scientifically valuable events for follow-up study while they remain bright.

\textbf{Multi-Modal Feature Extraction and Reconstruction:}
We consider two tasks in this work: 1) inferring the time-evolving spectroscopic behavior of a supernova from measurements of its light curve (photometry) alone, and 2) learning a joint data representation for photometry and spectra that is useful for downstream tasks. Though feature learning methods using contrastive learning \citep{zhang2024maven} and pure generation methods using diffusion \citep{shen2025variational} have been applied to these modalities, a single approach has not yet been presented that can simultaneously achieve goals 1 and 2. As an obstacle toward these goals, the correlations between light curves and spectra are many-to-many: a spectrum obtained at a single time can be associated with multiple feasible light curves, and each light curve is associated with multiple spectra describing the wavelength-dependent emission of an explosion at different times. Furthermore, the dimensionality of each modality varies with the total number of observations obtained and the resolution of the instrumentation used. For these reasons, our proposed method should be able to 1) learn compact private and shared representations, 2) operate on a variable number of modalities and input lengths at inference time, and 3) keep inference costs low as input dimensionality increases. 

We achieve these goals by constructing a Perceiver-IO \citep{jaegle2021perceiverio} style model for time-domain inputs, which we call the \texttt{transceiver} (for ``transient perceiver"). We combine this model with a mixture-of-expert VAE \citep[MMVAE,][]{shi2019variational} for private and shared feature learning, and validate its use for cross-modality generation. Our model and associated experiments can be found at the github repository for this work \footnote{\url{https://github.com/YunyiShen/VAESNe-dev}}.

\section{Method}
\label{sec:method}
The data used in this work are less structured than modalities typically encountered in machine learning tasks (e.g., text, images, and audio). We begin by describing our data and the chosen encoder-decoder architecture. Then, we introduce the architecture and training objective associated with our \texttt{MMVAE}. 

\textbf{Data Encoding:}
Our approach follows the encoding scheme of \citet{zhang2024maven}, but augments it with learnable positional encodings and phase information to enable multiple spectra to be linked to a single light curve. 

\textbf{Photometry:}
For observation $i$ obtained with photometric filter $n$ during an imaging survey containing $B$ unique photometric filters, $(t_{n,i}, m_{n,i}, b_{n,i}) \in \R\times \R \times [B]$ represents the observation time, filter, and measured magnitude of the explosion. Consequently, the full light curve is a sequence of $(t_{n,i}, m_{n,i}, b_{n_i})$ with length $L_{\text{photo}, n}$. 

Our light curves are encoded to a model dimension $d$ by summing a sinusoidal encoding for time, a class embedding of filter, and linear projection of magnitude, resulting in an embedded sequence of dimensionality $\R^{L_{\text{photo}, n}\times d}$ that was passed into perceiver. We encode observation times with a sinusoidal positional encoding followed by a learnable MLP \citep{peebles2023scalable}. Magnitude is encoded using a learnable linear embedding and the photometric filter is encoded using a simple categorical embedding. For parallelization, we also pad our input light curves to a consistent length; we retain a mask during training to indicate padded positions. 

\begin{figure*}[htp]
\centering
\includegraphics[width=0.9\linewidth]{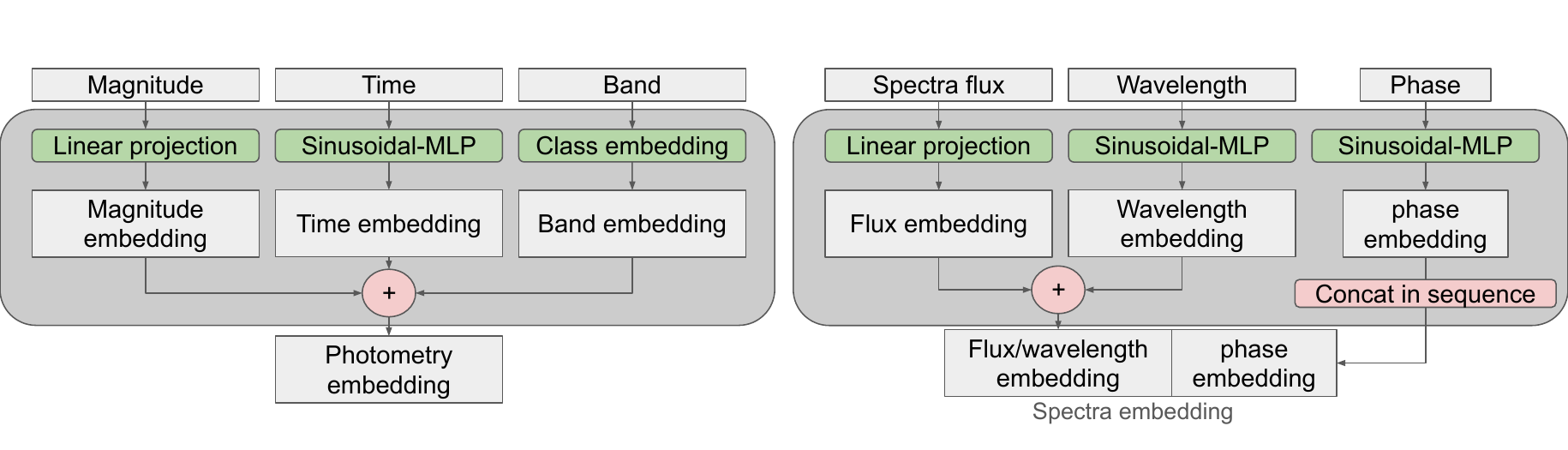}
\vspace{-0.6cm}
\caption{Embedding method used for supernova spectra and light curves (see text for details). The approach is similar to that used in \citet{zhang2024maven}, but with learnable positional encodings for the observation times of photometry and spectroscopy.}
\label{fig:modality}
\end{figure*}

We visualize the full architecture of our model in \cref{fig:modality}. 

\textbf{Spectroscopy:}
Spectra encode the flux from the supernova as a function of wavelength. Each spectrum is taken at a fixed point in time, which we represent as phase with respect to the time of supernova peak light. The $n$th spectrum associated with each supernova is represented as a sequence of flux values $f_{n,i}$ measured at wavelengths $\lambda_{n,i}$. As a result, the spectra for each supernova have dimensionality $L_{\text{spec}, n}$. 

We encode each spectrum by separately embedding wavelengths and fluxes to the model dimension $d$ and summing them. Wavelengths are encoded using sinusoidal embeddings followed by a learnable MLP, while flux values are embedded linearly. This results in an embedding of shape $\R^{L_{\text{spec}, n}\times d}$. Unlike photometry, we separately embed the phase using a sinusoidal embedding followed by an MLP projection to dimension $d$, and then append it to the sequence as a special token. This special token approach allows for more flexible information passing through cross-attention. The full spectrum is thus embedded into dimensionality $\R^{(L_{\text{spec}, n}+1)\times d}$. As with photometry, we pad input spectra and retain a mask indicating which positions are padded.

\textbf{\texttt{Transceiver} Encoder-Decoder:}
We use an architecture akin to Perceiver and Perceiver-IO \citep{jaegle2021perceiver, jaegle2021perceiverio} for our encoder-decoder setup. This architecture handles \textbf{any-length sequences} via cross-attention. Specifically, in the encoder, a fixed-size latent sequence (as query) attends to the input spectrum/light curve embedding (as key and value) to produce the latent posterior mean and variance. We refer to this architecture for transients as the \texttt{transceiver}. This differs from the vanilla transformer encoder used in models like \texttt{Maven} \citep{zhang2024maven}, in which the input data itself serve as query and therefore must be fixed-size.

\begin{figure}[htp]
\centering
\includegraphics[width=0.9\linewidth]{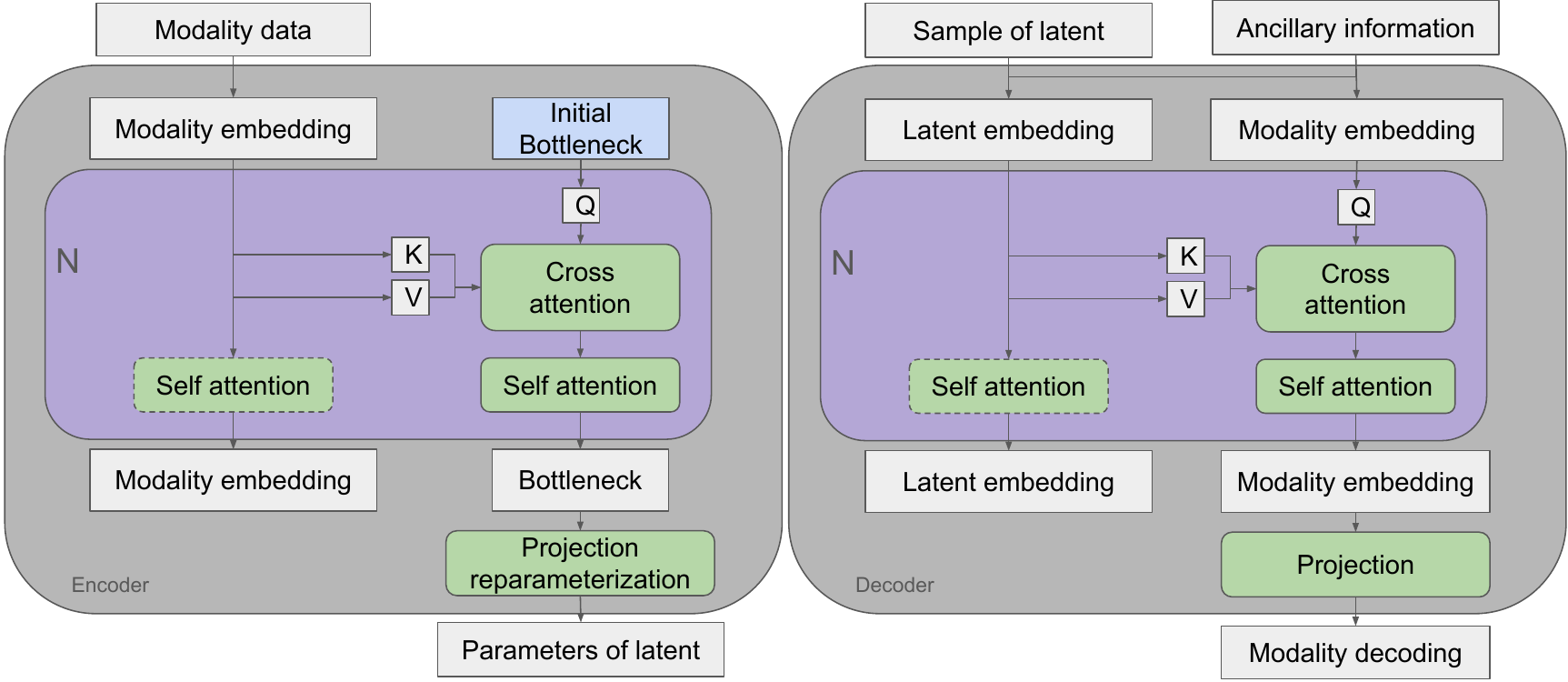}
\vspace{-0.3cm}
\caption{Encoder-decoder architecture with \texttt{perceiver-IO}-style architecture for individual modalities. Modality encodings are cross attended by a latent bottleneck sequence.}
\label{fig:encoderdecoder}
\vspace{-0.7cm}
\end{figure}

\textbf{Multi-Modal Mixture-of-Expert VAE:}
\citet{shi2019variational} proposed using a mixture of-expert VAE (\texttt{MMVAE}) to combine different modalities, enabling learning of both shared and private information. The key idea in \texttt{MMVAE} is to model the latent variable $z$ as a mixture of distributions associated with the different modalities. Formally, let $z_n$ denote the latent variable for the $n$th training pair and $x_{n,m}$ denote the data from modality $m=1, \dots, M$. Our \texttt{MMVAE} approximates the conditional latent distribution, denoted $q_{\phi_m}$, using \texttt{transceiver} encoders parameterized by $\phi_m$. Each encoder outputs the mean and variance of a Laplace distribution:
\begin{equation}
p(z_n|x_{n,1},\dots, x_{n,M})\approx q_{\phi}:= \frac{1}{M}\sum_{m=1}^M q_{\phi_m}(z_{n}|x_{n,m})
\end{equation}
Though learned jointly, we can use the encoder for each modality independently or combine them via the mixture-of-experts model at inference time. Each modality also has its own decoder. That is, with a decoder parameterized by $\psi_m$,
\begin{equation}
p(x_{n,1,\dots,M}|z_n)\approx\prod_{m=1}^M q_{\psi_m}(x_{n,m}|z_n)
\end{equation}

From this perspective, the model can be seen as a generalization of  \citet{2025ShenGagliano}, in which a diffusion model was constrained to producing spectroscopic posteriors conditioned on obtained photometry (though the two architectures are notably different).

We use the \texttt{transceiver} decoder to parameterize the mean of a Laplace distribution for each modality. We model the prior $p(z)$ as a standard Laplace distribution with latent space size $z\in \R^{4\times 4}$.
Training uses the IWAE objective proposed by \citet{shi2019variational}:
\begin{equation}
\mathcal{L}(x_{1,\dots, M})=\mathbb{E}_{z^{1:K}\sim q{\phi}}\left[\log\sum_{k=1}^K\frac{1}{K}\frac{p(z)\prod_{m}q_{\psi_m}}{q_{\phi}}\right]
\end{equation}
We train using \texttt{AdamW} in PyTorch using a learning rate of 0.001 for 500 iterations. For our \texttt{transceiver} encoder and decoder, we have 4 layers of multi-headed attention with 4 heads each. The model operates in the dimension of 32, with an MLP dimension of 32 in the attention layers. We choose a latent dimension of 4 by 4. 

\textbf{Uni-Modal VAE for Spectra Generation:}
To evaluate our model's capacity for cross-modality conditional generation, we build a reference baseline: a VAE trained directly on spectral data using the same \texttt{transceiver} encoder-decoder architecture but trained with the standard ELBO loss. This baseline ignores light curve information completely.

\textbf{Multi-Modal \texttt{Transceiver} Trained via Contrastive Learning:}
As a comparison, we also jointly train  \texttt{transceiver} encoders using a constrastive objective. We project light curves and spectra into a shared $\R^{4\times 4}$ latent space, further project them to $\R^{4\times 8}$ using a learned MLP, compute the Frobenius distance between pairs, and train the model to minimize the InfoNCE loss \citep{khosla2020supervised}.

\section{Experiments}
\label{sec:experiments}
\textbf{Data:}  
We use the simulation grid from \citet{goldstein2018evidence} to train and validate our \texttt{MMVAE}. \citet{goldstein2018evidence} generates 4,500 Type Ia supernovae (SNe~Ia) using the \texttt{Sedona} radiation transport code \citep{kasen2006secondary}. Each simulation yields a full spectral energy distribution (SED) surface, sampled at 1-day intervals in time and $\sim$30~\AA\ intervals in wavelength. We present the physical parameters associated with the simulation grid in Table~\ref{tab:sim_params} in Appendix~\ref{app:params}.

We simulate LSST-like photometry by integrating the SED surfaces across the transmission curves of the LSST filters, obtained from the SVO Filter Profile Service\footnote{\url{http://svo2.cab.inta-csic.es/}}. The light curve cadence is based on the baseline v3.3 simulations of LSST's Wide-Fast-Deep survey. We assume a random distribution of events in both time since survey start and position across the survey footprint.

After sampling photometry, we require at least 10 total photometric observations across all bands to include an event in our train/validation/test sets. For simulated spectra, we extract fluxes directly from the SED surface in 10-day windows from 10 days before peak brightness to 30 days after it. We apply a median filter with bandwidth 3 to approximate the finite resolution of spectrographs. We then train the model on the log-transformed flux values in units of $\mathrm{erg}\;\mathrm{s}^{-1}\;\mathrm{cm}^{-2}\;\mathrm{Hz}^{-1}$. All flux, wavelength, phase, and magnitude values are standardized to z-scores during training and converted back to physical units at inference time. Light curves are zero-padded to 10 measurements per band during training solely for parallelization, yielding uniform arrays of length 60. We split the dataset into fractions of 80/10/10 for training, validation, and testing, respectively.

\textbf{Prior Sampling:} To evaluate the spectroscopic diversity learned by the model, we sample from both the latent (standard Laplace) prior and the conditionally-generated spectra from events in our test set. We present the results in \cref{app:prior}. The samples resemble the observed diversity in the test sample with some smoothing of high-frequency information, as is commonly observed in VAEs. 

\textbf{Cross-Modality Generation:}  
In this setup, we encode photometry using the trained encoder, sample from the posterior distribution in latent space, and decode into spectra using the trained decoder. We evaluate our conditional generation using three metrics: the residual error in the posterior mean, the credible interval (CI) coverage, and the CI width. As a baseline, we compare to (i) a uni-modal VAE trained to encode spectra alone, (ii) a nearest neighbor search in a learned contrastive latent space\footnote{The nearest neighbor retrieval does not provide uncertainty estimates, so we exclude it from our CI-based comparisons.}, and (iii) the average spectrum across the full training set at each phase. The first benchmark allows us to confirm that our model architecture has sufficient capacity to reconstruct each spectrum, while the third ensures that our reconstructions are successfully conditioned by the event's photometry.

\begin{figure*}[htp]
    \centering
    \vspace{-0.35cm}
    \includegraphics[width=\linewidth]{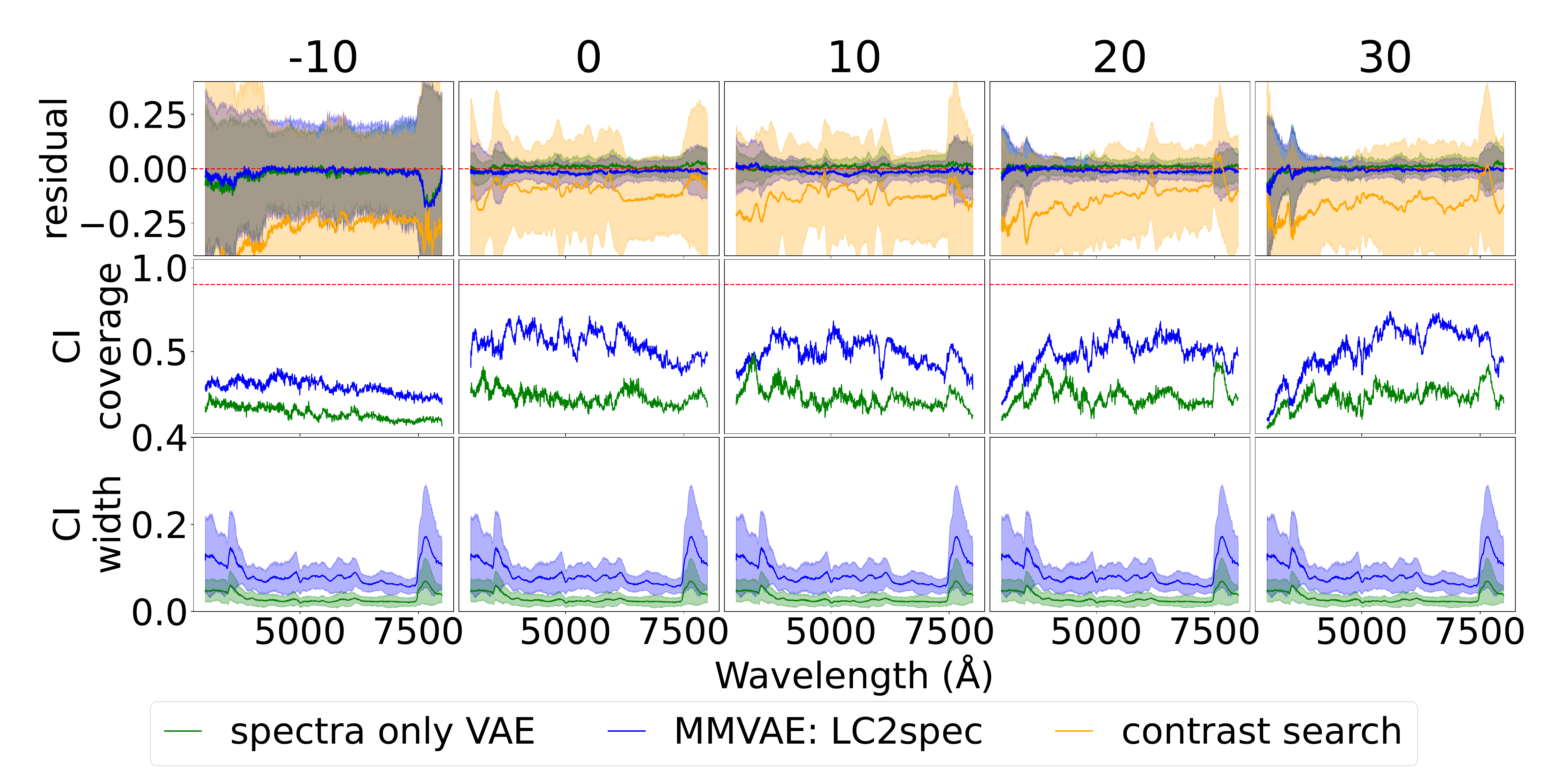}
    \vspace{-0.8cm}
    \caption{Evaluation metrics for the generation of type Ia supernova spectra conditioned on LSST light curves: the difference between the predicted and ground truth spectrum (top row); the CI coverage (middle row) and CI width (bottom row). Spectral phases relative to SN peak light are given at top in days. Our \texttt{MMVAE} (in blue) achieves similar reconstruction accuracy to a model trained only on spectra (green), and outperforms a nearest neighbor search in the contrastive latent space (orange line). Unsurprisingly, the spectra-only VAE However, the low CI coverage indicates that the posterior samples are not well-calibrated.
    }
    \vspace{-0.2cm}
    \label{fig:reconmetric}
\end{figure*}

Our results are shown in \cref{fig:reconmetric}. Our \texttt{MMVAE} achieves reconstruction performance comparable to the spectra-only VAE and exceeding that of the contrastive method, which essentially retrieves the average SN~Ia spectrum at each phase. Further, the model's performance remains robust under increased masking of light curve inputs (see \cref{app:masking} for details).The posterior samples of the model, however, are not well-calibrated, as the CIs exhibit undercoverage. This may reflect the limited flexibility of the encoder in approximating complex posterior distributions. 

\begin{figure*}[!h]
    \centering
    \includegraphics[width=\linewidth]{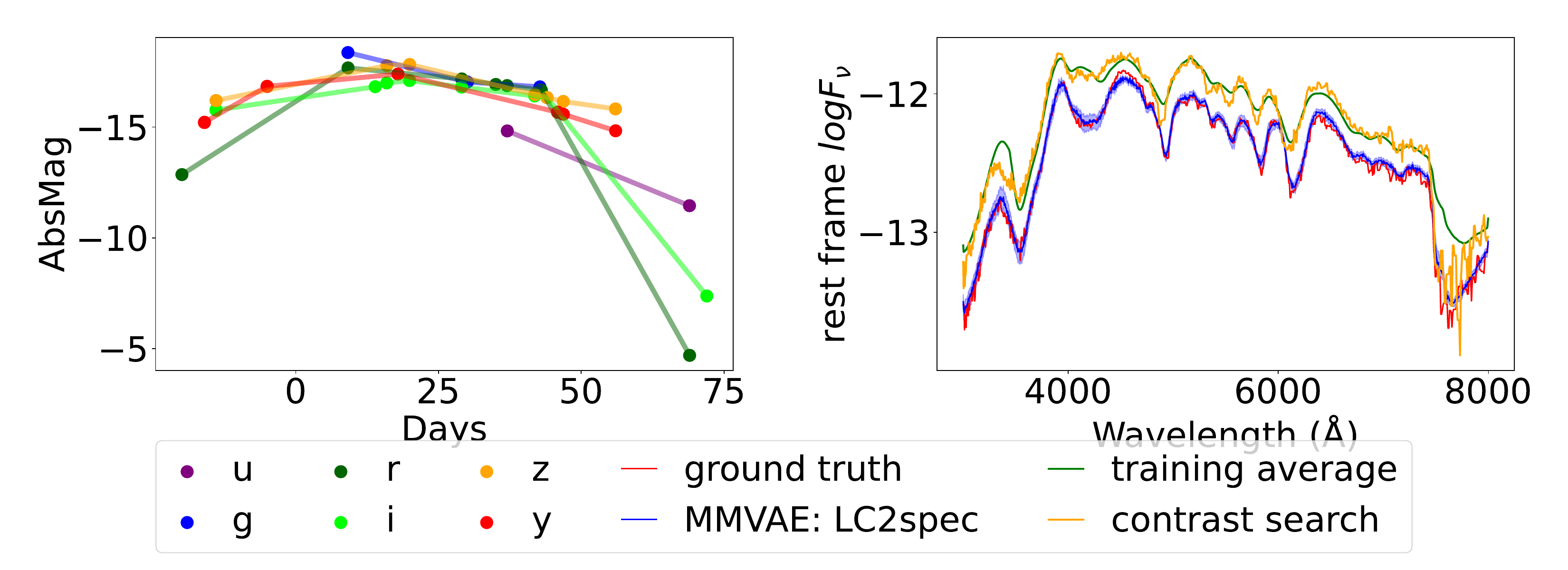}
    \vspace{-0.8cm}
    \caption{Sample reconstruction of a spectrum 10 days after supernova peak light by our \texttt{MMVAE} (right, `LC2spec'), and conditioned on a simulated LSST light curve in six filters (left). Baseline spectra are shown at right for the training average at +10 days in green and the nearest spectrum recovered by the contrastive model in orange (see text for details).}
    \label{fig:examplereconstruct}
\end{figure*}

We show a representative example of a spectrum decoded by our \texttt{MMVAE} and the baseline models in \cref{fig:examplereconstruct}.
 
\textbf{Feature Extraction:}  
In this task, we evaluate whether sparsely-sampled light curves can be used to recover the underlying physical parameters from the Goldstein simulation grid. We encode each light curve and fit a series of small MLPs to regress the simulation parameters in Table~\ref{tab:sim_params}. As a baseline, we train an end-to-end \texttt{transceiver} encoder to regress the model parameters directly from both photometry and spectra. Our results can be found in \cref{app:LC2goldstein}.

\section{Discussion}
\label{sec:discussion}
We have introduced an \texttt{MMVAE} trained to learn joint embeddings between supernova modalities and perform probabilistic cross-modality generation. Our proposed \texttt{transceiver} encoder and decoder has independent value beyond the \texttt{MMVAE}, and can be used to train alternative multi-modal architectures like flamingo \citep{alayrac2022flamingo} for representation learning of time-domain datasets. 

The model presented can be used to inform real-time follow-up of supernovae. By producing posteriors over spectra conditioned on photometry, it allows astronomers to rapidly infer an explosion's general physical properties and to triage follow-up targets -- either to confirm predicted spectral features or collect data to improve future predictions. 

Beyond reconstruction, the model's latent features can also serve as inputs to lightweight, task-specific models. Empirically, our \texttt{MMVAE} can reconstruct data modalities more faithfully than a similar model trained with a constrastive objective. As a result, our approach provides a promising alternative for training multi-modal ``foundation models" in astrophysics \citep[e.g.,][]{parker2024astroclip,zhang2024maven,rizhko2025astrom3}. The intermediate cross-modal reconstructions can also be used for mechanistic interpretability studies, by linking predicted spectral features to the output of downstream models optimized for classification and redshift prediction.

In future work, we will incorporate host-galaxy images of the observed supernova as a third modality and extend our architecture to explicitly enforce public and private information using, e.g., \texttt{MMVAE}+ \citep{palumbo2023mmvae+, martens2024disentangling} or diffusion decoders \citep{palumbo2024deep}. We also aim to improve the calibration of our output samples. It is possible that the strong performance of the model reflects the low intrinsic diversity among SNe~Ia compared to, e.g., core-collapse supernovae; we plan to re-train the model on synthetic grids of more heterogeneous classes to further validate the technique. Finally, we will fine-tune the model for common downstream tasks in transient science and benchmark performance against existing models (e.g., ParSNIP for classification, \citealt{2021Boone_Parsnip}; SALT3 for joint fitting of photometry and spectra, \citealt{2021Kenworthy_SALT3}).

\section*{Acknowledgements}
We are grateful to the anonymous reviewers of the ICML-colocated ML4Astro
2025 workshop for their constructive feedback, and to Siddharth Mishra-Sharma for early discussions that shaped this work. This work is supported by the National Science Foundation under Cooperative Agreement PHY-2019786 (The NSF AI Institute for Artificial Intelligence and Fundamental Interactions, http://iaifi.org/).

\section*{Impact Statement}
The proposed method can be used to learn joint representations between multiple modalities with a probabilistic formalism. The method can be generalized to other datasets with covariate modalities. 
\nocite{langley00}

\bibliography{example_paper}
\bibliographystyle{icml2025}

\newpage
\appendix
\onecolumn
\section{Simulation Parameters}
\label{app:params}

We present the physical parameters associated with the \citet{goldstein2018evidence} simulation grid in \cref{tab:sim_params} below.

\begin{table}[H]
    \centering
    \caption{Physical parameters associated with the SN~Ia simulation grid used in this work.}
    \begin{tabular}{lcc}
        \hline\hline
        Parameter & Symbol & Range \\ \hline
        Kinetic energy            & $E_k$              & $[4.514,\,5.144]\times10^{51}\,\mathrm{erg}$ \\
        Ejecta Mass 
                                  & $m_{\mathrm{ej}}$  & $[0.7,\,2.5]\,M_\odot$                       \\
        $^{56}$Ni Mass            & $m_{\mathrm{Ni}}$  & $[0.75,\,2.16]\,M_\odot$                    \\
        IME Mass 
                                  & $m_{\mathrm{IME}}$ & $[0.0036,\,2.07]\,M_\odot$                  \\
        Unburnt C/O Mass          & $m_{\mathrm{CO}}$  & $[0,\,0.17]\,M_\odot$                       \\
        Progenitor Mass           & $m$                & $[1.0,\,2.5]\,M_\odot$                      \\ \hline
    \end{tabular}
    \label{tab:sim_params}
\end{table}

\section{Additional Results}
\subsection{Prior sampling}
\label{app:prior}
\begin{figure}[htp]
    \centering
    \includegraphics[width=\linewidth]{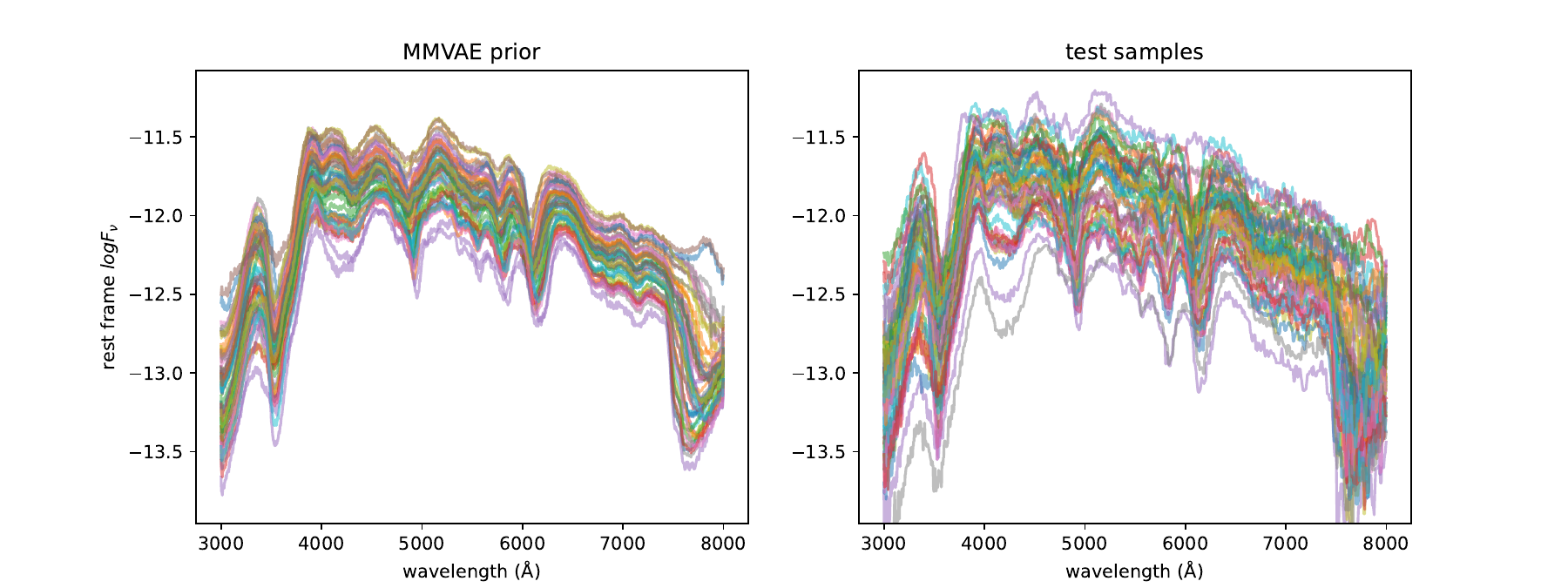}
    \caption{Spectral samples from the model prior (left) and after conditioning on photometry for events in the test set (right). Our model captures key spectroscopic diversity in SNe~Ia e.g., the blended absorption near 4200~Å, and this diversity is enhanced by the inclusion of cross-modality information.}
    \label{fig:priorspectra}
\end{figure}

We can sample the latent prior and decode directly into spectra to inspect the model's diversity in generation. We show 50 prior samples of spectra at +10 days relative to peak light in \cref{fig:priorspectra}. The model's prior qualitatively matches the test sample but unsurprisingly retains less high-frequency structure.

\subsection{Masking}
\label{app:masking}
\begin{sidewaysfigure}[htp]
    \centering
    \includegraphics[width=\linewidth]{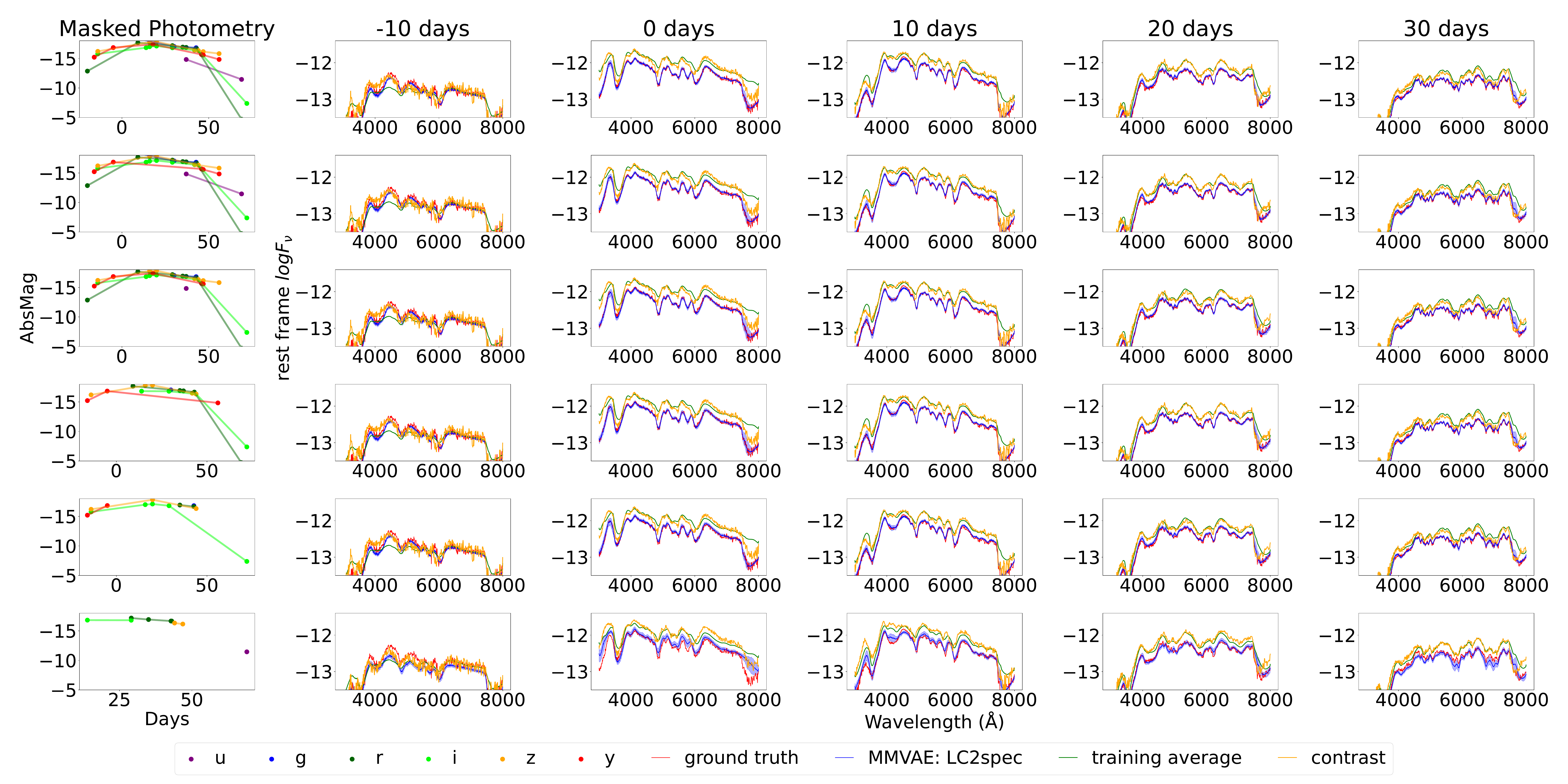}
    \caption{Sample spectra reconstructions (right columns) obtained by conditioning on photometry at different levels of masking (left column). Rows correspond to 0, 10, 30, 50, 70, and 90\% masking fractions and right columns correspond to spectra at phases of -10, 0, +10, +20, and +30 days relative to peak luminosity. Right panels show ground truth spectra (red), \texttt{MMVAE}-generated spectra (blue), the average SN~Ia spectrum in the training set at each phase (green), and the nearest-neighbor spectra from the contrastive model (orange). Spectra are shown in units of $\mathrm{erg}\;\mathrm{s}^{-1}\;\mathrm{cm}^{-2}\;\mathrm{Hz}^{-1}$ and photometry is shown in units of absolute magnitudes.}
    \label{fig:masking}
\end{sidewaysfigure}

In this experiment, we randomly mask 0\%, 10\%, 30\%, 50\%, 70\%, and 90\% of the conditioned light curves and attempt to reconstruct a spectrum. An example is shown in \cref{fig:masking}. Our method shows some robustness to masking until nearly 70\% of photometric observations are masked, beyond which model predictions appear to collapse to the training-set mean spectrum. 

\begin{sidewaysfigure}[htp]
    \centering
    \includegraphics[width=\linewidth]{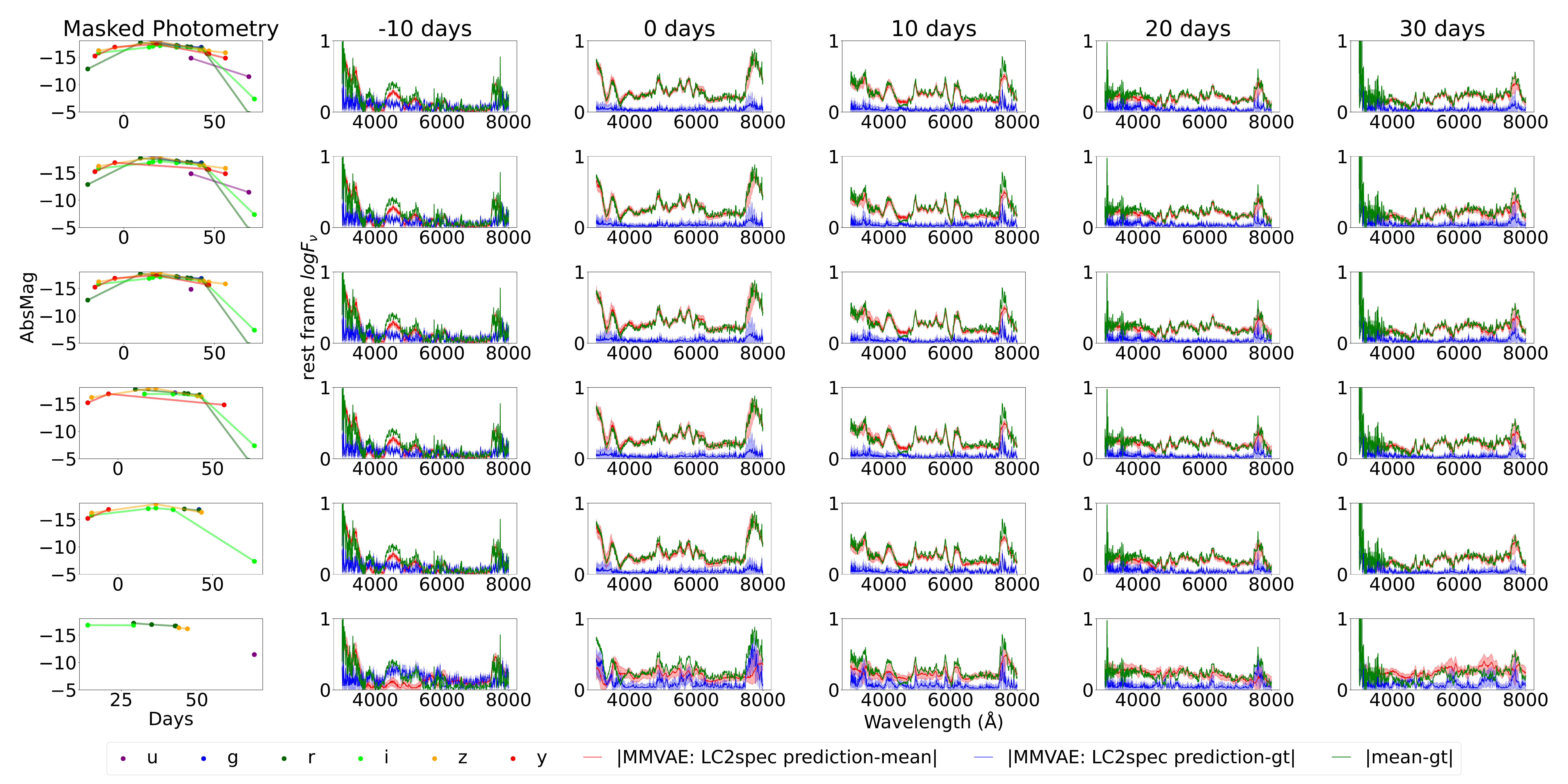}
    \caption{Light curves (left column) and absolute residuals (right columns) between the predicted and mean SN~Ia spectrum (red), predicted and ground-truth event spectrum (blue), and mean SN~Ia and ground-truth event spectrum (green). We observe a decrease in $|$LC2spec-mean$|$ at all wavelengths as more photometry is masked, confirming that our \texttt{MMVAE} regresses toward the mean prediction when conditioned on fewer observations. This phenomenon is most severe when predicting spectra prior to peak light.}
    \label{fig:markingresidual}
\end{sidewaysfigure}

Next, we more directly assess whether our model reverts to predicting an average SN~Ia spectrum when conditioned on less photometric information. In \cref{fig:markingresidual}, we show the absolute difference between 1) the model prediction and an average SN~Ia spectrum from our grid, 2) the model prediction and the ground-truth spectra, and 3) the average and ground-truth spectra. We observe that as the majority of observations are masked, the difference between predicted and average spectra systematically decreases. This confirms that the model regresses toward the prior when conditioned on less information. 

\newpage
\subsection{Feature extraction}
\label{app:LC2goldstein}
The baseline \texttt{end2end} model is a simple encoder following the 
\texttt{transceiver} architecture, similar to the encoder used in \texttt{MMVAE} and the contrastively-trained model.

\begin{figure}[htp]
    \centering
    \includegraphics[width=0.9\linewidth]{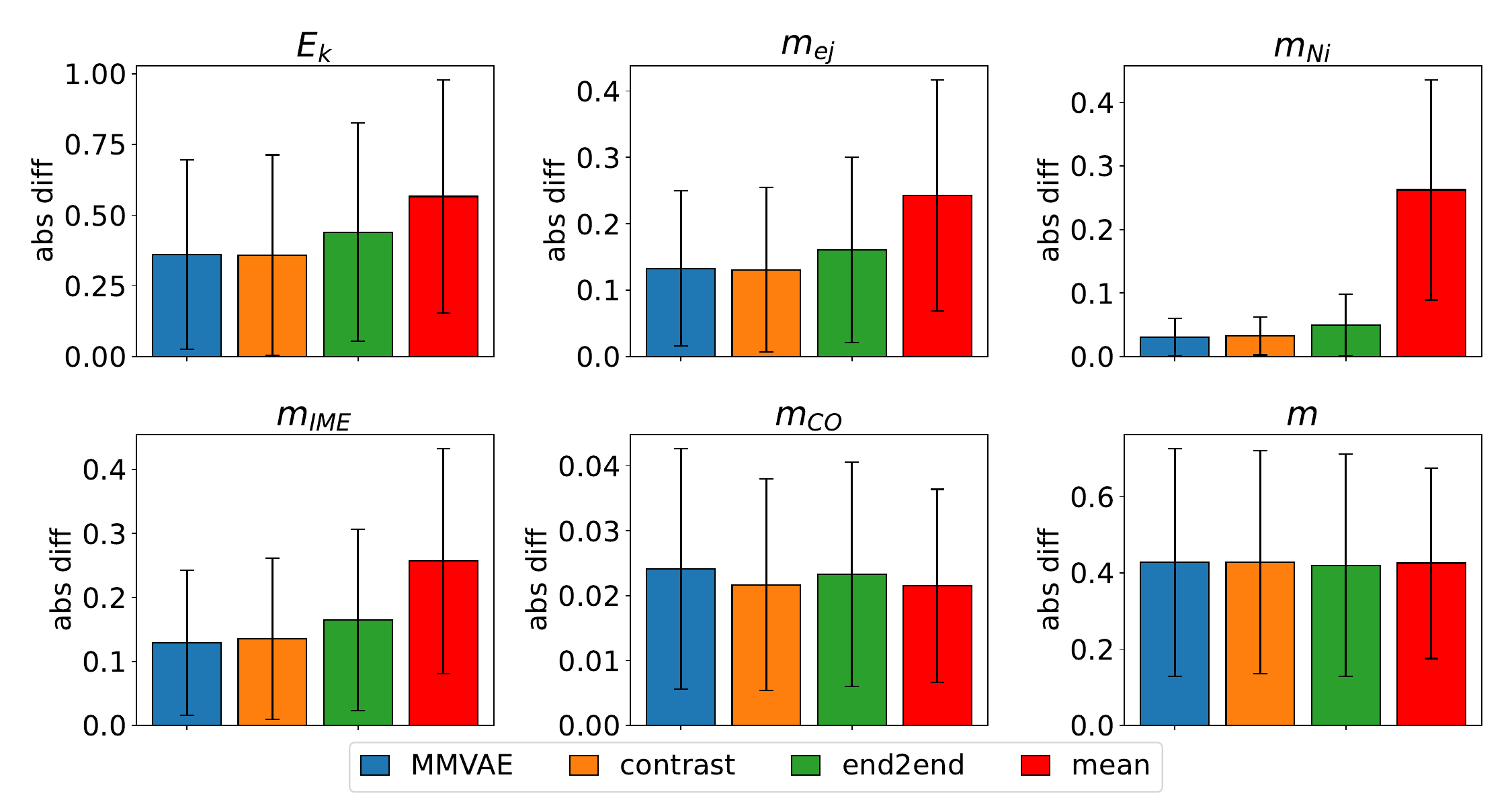}
    \vspace{-0.5cm}
    \caption{Mean and 1$\sigma$ standard deviations of absolute residuals for recovering six parameters from \citet{goldstein2018evidence}'s simulation grid. Both \texttt{MMVAE} and contrastive models outperform end-to-end regression on most parameters, but with large scatter. 
    }
    \vspace{-0.5cm}
    \label{fig:LC2goldstein}
\end{figure}

We train identical MLPs to predict each physical parameter in Table~\ref{tab:sim_params} using two-thirds of the test set. We evaluate on the remaining third. Our results are shown in \cref{fig:LC2goldstein}.

Recovering most physical parameters from photometry alone proves challenging, and most methods perform only slightly better than a training-set average. This may be the result of the sparse sampling of the conditioned LSST light curves. Parameters closely tied to brightness ---such as $^{56}$Ni mass --- are better predicted. Both the \texttt{MMVAE} and the contrastively-trained model outperform the end-to-end baseline on most parameters, suggesting that a structured latent representation is better able to preserve physical information. We aim to explore this hypothesis with additional experiments in future work.

\end{document}